\documentstyle[12pt]{article}
\renewcommand{\S}{\hat S}
\renewcommand{\a}{\dot a}
\newcommand \bq {\begin{eqnarray}}
\newcommand \be {\begin{equation}}
\newcommand \eq {\end{eqnarray}}
\newcommand \ee {\end{equation}}
\newcommand \pr {\prime}
\newcommand \RE {{\rm Re}}
\newcommand \IM {{\rm Im}}
\newcommand \half {\frac{1}{2}}
\newcommand \quart {\frac{1}{4}}
\newcommand \ran {\rangle}
\newcommand \lan {\langle}
\newcommand \om {\omega}
\newcommand \al {\alpha}
\newcommand \si {\sigma }
\newcommand \bt {\beta}
\newcommand \ep {\epsilon}
\newcommand \ga {\gamma}

\newcommand \la {\lambda}
\newcommand \r {\rho}

\begin{document}

\setcounter{page}{1}
\vspace{2cm}
\begin{center}
\vspace{5mm}
\vspace{5mm} {\large Multi-access channels in quantum 
information theory.}\\
\vspace{5mm} {\large A.E. Allahverdyan, D.B. Saakian}\\
\vspace{5mm}
{\em Yerevan Physics Institute}\\
\end{center}

\vspace{5mm}
\centerline{\bf{Abstract}}
The multi-access channels in quantum information theory are considered.
Classical messages from independent sources,
which are represented as some quantum states, are transported
by a channel to one address. The messages can interact with each other
and with external environment.
After statement of problem and proving some general results
we investigate physically important case when information is
transported by states of electromagnetic field.
One-way communication by noisy quantum channels is also considered.

\section{Introduction} 
\vspace{5mm}
Physical ideas played important role as sources 
for information theory
\cite{cavesdr,strat}.
Only in 15 years after discovery of information theory 
in works of Gordon, Lebedev and Levitin, and other researcher (for references see \cite{cavesdr})
the general structure of information theory has been considered from
physical point of view.
Indeed quantum-mechanical and thermodynamical limitations
are very important for correct consideration of information-theoretical
models \cite{cavesdr,strat}.\\
More generally, quantum information theory contains two
distinct types of problem. The first 
type describes 
transmission of classical information through a 
quantum channel 
(the channel can be noisy or noiseless).
In such scheme bits encoded as 
some quantum states and only this states or its
tensor products are transmitted. 
In the second case arbitrary 
superposition of this states or entanglement 
states are transmitted.
In the first case the problems can 
be solved by methods of classical information theory,
but in the second case new physical representations are needed.
In this work we investigate the problems of the first type.\\
In the almost all paper devoted to problems of physical information theory only one-way 
communication are considered- i.e., there is one input with some initial quantum ensemble and one
output where quantum states are detected. Here quantum ensemble is some set
of quantum states (which can be represented 
by corresponding density  matrices ) with corresponding probabilities. 
But in the practice  multi-terminal communication schemes also are important. In this 
case there are several input for information and several output for detection. 
In this communication scheme   messages are represented as 
some physical systems and there are  interactions between the messages 
and external environment.
For general discussion about multi-terminal classical information theory see \cite{coverelgammal}.\\
Now about  the concrete statement of problem.
We can consider one output which receive information from two independent sources.
The messages of these sources are represented as some quantum states. More exactly we can say  that
for any letter of some classical alphabet the concrete quantum state is generated.
The quantum ensembles of two independent sources are
\begin{equation}
\label{1}
\rho ^{(1)} =\sum_{\alpha }p^{(1)}_{\alpha }\rho ^{(1)}_{\alpha
},
\end{equation}
\begin{equation}
\label{2}
\rho^{(2)} =\sum_{\beta }p^{(2)}_{\beta }\rho ^{(2)}_{\beta }.
\end{equation}
After initial preparation the quantum states of (\ref{1}, \ref{2}) penetrate through a 
quantum channel. In this channel there are interactions between the states of  sources
and an interaction with the environment.
The concrete mechanisms of this interactions will be discussed later.
The general effect of the noisy quantum channel can be described by a 
quantum evolution operator 
$\hat {S}$ with kraussian representation
\begin{equation}
\label{4}\hat S\rho =\sum_\mu A_\mu ^{\dag }
\rho A_\mu ,\ \ \sum_\mu A_\mu
A_\mu ^{\dag }=\hat 1.
\end{equation}
This operators must be linear, completely
positive and trace-preserving \cite{krauss,shum}. 
After interaction receiver has the states $\si ^{(12)}_{\al \beta }$
\begin{equation}
\label{5}
\S (\rho ^{(1)}_{\alpha }\otimes \rho
^{(2)}_{\beta })=\sigma ^{(12)}_{\alpha \beta }
\end{equation}
These are states of quantum ensemble 
\begin{equation}
\label{dopdop1}
\si =\sum_{\al ,\bt }p_{\al }p_{\bt }\si ^{(12)}_{\al \bt}
\end{equation}
At the output of the channel receiver should separate and recognized the messages of sources.
We can say that the receiver  need some measurement procedure.
It is important that the elements of (\ref{1},\ref{2}) can be nonorthogonal or 
nonorthogonality can occur after action of (\ref{5}).
In this case for more optimal distinguishing between different
quantum states we need generalized measurement procedure \cite{helstrom}.
This type of measurement is represented by some nonorthogonal 
resolution of identity
\bq
\label{6}
&&\sum_{\ga }E_{\ga}=1 \nonumber \\
&&E_{\ga } > 0,\  \ E_{\ga }=E_{\ga }^{\dag }
\eq
If system with density matrix $\rho $ is measured
then probability of result $\ga $ is ${\rm tr}(\rho E_{\ga })$.
In the \cite{holevonova} was shown that for 
distinguishing nonorthogonal states measurements like (\ref{6})
more optimal than usual.
In the output of the channel (\ref{5}) as result of some 
generalized measurement like (\ref{6}) we have
the conditional probabilities 
\begin{equation}
\label{7}
p(\gamma /\alpha \beta )={\rm tr}(E_{\gamma }
\sigma ^{(12)}_{\alpha \beta })
\end{equation}
This procedure is called decoding.
With initial distributions 
\be
\label{8}
 p(\al),\   \
 p(\bt )
\ee
(\ref{7}) determined the definition of the binary multi-access channel in the 
classical sense \cite{coverelgammal,strat}.
Let $R_{1}$, $R_{2}$ are the maximal quantities of  information that the
sources can transport in the regime of the reliable
connection (it is the connection with small probability of error in the decoding).
As was shown in the \cite{coverelgammal}
such $R_{1}$, $R_{2}$ must satisfied the following conditions
\begin{equation}
\label{10}
R_1< I(\alpha :\gamma /\beta ),\  \
R_2< I(\beta :\gamma /\alpha ),\  \
\end{equation}
\begin{equation}
\label{11}
R_1+R_2< I(\alpha \otimes \beta :\gamma ).
\end{equation}
Where
\begin{equation}
\label{24}
I(\alpha :\gamma /\beta )=\sum_{\alpha ,\beta ,\gamma }
p(\alpha ,\beta ,\gamma )\ln \frac{p(\gamma /\alpha \beta
)}{p(\gamma /\beta )},
\end{equation}
\begin{equation}
\label{24a}
I(\alpha \otimes \bt :\gamma )=\sum_{\alpha ,\beta ,\gamma }
p(\alpha ,\beta ,\gamma )\ln \frac{p(\gamma /\alpha \beta
)}{p(\gamma )}.
\end{equation}
The second value is usual mutual information between ensembles $\gamma $ and $\al \otimes \bt$.
The first  value is called mutual-conditional information
(mc-information). The mutual information of
two ensembles is reduction of entropy of
one ensemble if the second is observed.
Mc-information has same meaning but after realization of
the conditional ensemble.
Obvious, that usual and famous formula of Shannon is particular
case of the (\ref{10}, \ref{11}).\\
The problem of physical information theory in this case 
is investigation the
results of  (\ref{10}), (\ref{11})
for physical important noisy channels.\\
The restriction (\ref{11})
is indeed important because for independent initial distribution we have
\begin{equation}
\label{ret1}
I(\al :\ga /\bt ) +I(\bt :\ga /\al )\geq I(\al \otimes \bt /\ga )
\end{equation}
In the general case the values (\ref{24}), (\ref{24a})
can not be calculated explicitly. Therefore
investigation the upper bounds for this values
is very important problem.
For the one-way communication such theorems was 
proved by A.Holevo \cite{holevodokvoteor}.
The most general results in this direction was
obtained in the \cite{yuenozawa}.
We shall obtain the  measurement independent 
upper bounds for 
(\ref{24}), (\ref{24a}) by method of \cite{yuenozawa}.
For this purpose we need some general
results from quantum statistical physics.\\
Quantum relative entropy between two
density matrices $\rho _1$, $\rho _2$ is defined as follows 
\begin{equation}
\label{st9}
S(\rho _1||\rho _2)={\rm tr}(\rho _1\log \rho _1-\rho _1\log \rho _2).
\end{equation}
This positive quantity was introduced by Umegaki \cite{umegaki}
and characterizes the degree of
'closeness' of density matrices $\rho _1$, $\rho _2$. The properties of
quantum relative information were reviewed by M.Ohya \cite{ohya}.
Here
only one basic property  is mentioned.
\begin{equation}
\label{st10}
S(\rho _1||\rho _2)\geq S(\hat S\rho _1||\hat S\rho _2).
\end{equation}
The  inequality was proved 
by Lindblad \cite{lindblad}. \\
In the derivations of our theorems
we use the method of the work \cite{yuenozawa}.
We have for mc-information
\begin{equation}
\label{l25}
I(\alpha :\gamma /\beta )=\sum_{\alpha ,\beta ,\gamma }
p^{(1)}_{\alpha }p^{(2)}_{\beta }
{\rm tr}(E_{\gamma }
\S (\rho ^{(1)}_{\alpha }\otimes \rho ^{(2)}_{\beta }))
\ln \frac{{\rm tr}(E_{\gamma }
\S (\rho ^{(1)}_{\alpha }\otimes \rho ^{(2)}_{\beta }))}
{{\rm tr}(E_{\gamma }
\S (\rho ^{(1)}\otimes \rho ^{(2)}_{\beta }))}
\end{equation}
For any generalized measurement
$I_{\gamma }$
and density matrix $\rho $
we define the following transformation
\begin{equation}
\label{l26}
\rho \mapsto \{{\rm tr}(I_{\gamma }\rho )\}\gamma .
\end{equation}
In other words $\rho $ is transformed
to  diagonal matrix with corresponding diagonal elements.
This map is also general quantum evolution operator like
(\ref{4}). Now we shall use map (\ref{l26}) and theorem (\ref{st10})
for values like
\begin{equation}
\label{l27}
{\rm tr}[
\S (\rho ^{(1)}_{\al }\otimes \rho ^{(2)}_{\bt })
(\ln \S (\rho ^{(1)}_{\al }\otimes \rho ^{(2)}_{\bt })
-\ln \S (\rho ^{(1)}\otimes \rho ^{(2)}_{\bt }))]
\end{equation}
We  get
\begin{equation}
\label{l28}
I(\alpha :\gamma /\beta )\leq
-\sum_{\alpha ,\beta  }
p^{(1)}_{\alpha }p^{(2)}_{\beta }
S(\S (\rho ^{(1)}_{\alpha }\otimes \rho ^{(2)}_{\beta }))
+\sum_{\beta  }
p^{(2)}_{\beta }
S(\S (\rho ^{(1)}\otimes \rho ^{(2)}_{\beta }))
\end{equation}
\begin{equation}
\label{l29}
I(\alpha \otimes \beta :\gamma )\leq
-\sum_{\alpha ,\beta  }
p^{(1)}_{\alpha }p^{(2)}_{\beta }
S(\S (\rho ^{(1)}_{\alpha }\otimes \rho ^{(2)}_{\beta }))+
S(\S (\rho ^{(1)}\otimes \rho ^{(2)}))
\end{equation}
These inequalities are very useful if general limits
are developed.\\
Now about information transmission by electromagnetic field.\\
In practice the most important tool for information transmission is
electromagnetic field.
We briefly recall the connection between formalism which is
described above and characteristics of an electromagnetic field
in the vacuum or liner dielectric media \cite{cavesdr}.
Here information is connected with longitudinal characteristics
of plane wave with fixed center-frequency and narrow bandwidth.
But transverse (polarization, wave vector) characteristics are fixed.
This statement of question is more or less
realizable in the practice.
Now about concrete statement of the problem. Two modes
of the
electromagnetic field with frequencies $\om _{1}$, $\om _{2}$
penetrate into  the noisy channel.
W have considered two kinds of  communication channels, which are coherent channel
and quadrature-squeezed (briefly squeezed ) channel. In the first case the inputs of the sources are coherent states
of chosen field modes, and information is carried in the pattern of complex amplitude excitations.
In the second case the inputs are squeezed states, and information 
is carried in the pattern of excitations of squeezed quadratures.
Relative to a coherent states a quadrature-squeezed states have reduced quantum uncertainty in 
one quadrature component, called the squeezed quadrature. There is a corresponding increase in the
uncertainty in the orthogonal quadrature component, called the amplified quadrature.
For the case of coherent states information can be recovered by heterodyne detection, i.e., by measuring both
quadratures of the mode. For the case of quadrature-squeezed states information is recovered by 
measuring of squeezed quadrature.
In the channel the modes  can interact together and with
external thermostat.
We assume that interaction between modes and interaction with the
thermostat are linear (for discussion about realization of this
type of interaction see \cite{luis}).

We describe our model by quantum Langevin equation \cite{lax}
where rotating wave approximation is done.
\begin{equation}
\label{f5}
i\a _1=\om _1 a_1-i\frac{\ga a_1}{2}+ka_2+i\bar{F}_1,
\end{equation}
\begin{equation}
\label{f6}
i\a _2=\om _2 a_2-i\frac{\ga a_2}{2}+ka_1+i\bar{F}_2
\end{equation}
Where $a_1$, $a_2$ are annihilation operators for the modes, $k$
is the strength of cross mode interaction, $\ga $ is the damping
constant (for simplicity we choose damping constant same for the
modes), $\bar{F}_1(t)$, $\bar{F}_2(t)$ are langevin forces
(white noise).
These equations are generated by Hamiltonian
\begin{equation}
\label{pom1}
H=\om _1a^{\dag }_1a_1+\om _2a^{\dag }_2a_2+k(a^{\dag }_1a_2+a^{\dag }_2a_1)
\end{equation}
Solution of these equations can be obtained immediately.
For example
\bq
\label{f7}
&&a_1=a_1(0)
(\ep e^{i\phi _1t}+(1-\ep )e^{i\phi _2t})
-a_2(0)\sqrt{\ep (1-\ep )}
(e^{i\phi _1t}-e^{i\phi _2t})\nonumber \\
   &&+\sqrt{\ep }e^{i\phi _1t}\int_{0}^{t}
                 e^{i\phi _1t^{\pr }}F_{1}(t^{\pr })dt^{\pr }
    +\sqrt{1-\ep }e^{i\phi _2t}\int_{0}^{t}
                e^{i\phi _2t^{\pr }} F_{2}(t^{\pr })dt^{\pr }
\eq
Where:
\be
\label{f9}
\phi _{1,2}=-\la _{1,2}+i\frac{\ga }{2} ,\ \  \la _{1,2}=\frac{\om _1+\om _2\pm 
\sqrt{(\om _1-\om _2)^2+4k^2}}{2}
\ee
\be
\label{f11}
\frac{1-\ep }{\ep }=\frac{(\la _1-\om _1)^2}{k^2}
\ee
\bq
\label{f13}
&&\langle F_{i}^{\dag }(t)
F_{i}(t^{\pr })\rangle =
\ga \bar{n}_T(\la _{1,2})\delta _{ij}\delta (t-t^{\pr })
\nonumber \\
&&\bar{n}_T=(\exp{(\la _{1,2}/T)}-1)^{-1}
\eq
\be
\label{f13a}
\langle F_{i}(t)
F^{\dag }_{j}(t^{\pr })\rangle =
\ga (\bar{n}_T(\la _{1,2})+1)\delta _{ij}\delta (t-t^{\pr })
\ee
\be
\label{f14}
\langle F_{i}F_{j}\rangle =0
\ee
Now we assume that modes $a_1$, $a_2$
at the $t=0$ (in the input of the channel)
are in squeezed states with squeezing parameters $r_1$, $r_2$
(some of these parameters can be zero) and shifts
($\al _1$, $\al _2$), ($\bt _1$, $\bt _2$)
(for definition of squeezed states see \cite{cavesdr} ).
In the output of the channel (at the moment $t$)
one of the modes (for example $a_1$) is measured.
The case when two modes are measured and decoders can interchange
the available information can be obtained by simple modification
of the final results.
We shall consider only two type of measurement.
In the case of heterodyning both component of the mode
is measured simultaneously.
In other words we have
\cite{cavesdr}
\be
\label{f18}
p(\ga /\al \bt )=\frac{1}{\pi }
\lan \ga |\r (t;a_1)|\ga \ran
\ee
Where  $|\ga \ran $ is coherent state with shift $\ga $,
and  $\r (t;a_1)$ the density matrix of the mode $a_1$
at the moment $t$.\\
In the second case one component of the mode is measured.
In this case
\be
\label{f19}
p(\ga _1 /\al \bt )=\lan \ga _1|\r (t;a_1)|\ga _1\ran
\ee
Where $|\ga _1\ran $ is eigenstate of the measuring component.
Later we shall treat homodyne and heterodyne measurements simultaneously, and
shall use symbol $\ga $ for the both cases.\\
Simple analysis shows  \cite{coverelgammal}
that capacities are maximized by gaussian input probabilities
\bq
\label{f20}
&&p(\al )\sim \exp{(-\frac{1}{2}\al ^TK_{\al }\al )}  \nonumber
\\
&&p(\bt )\sim \exp{(-\frac{1}{2}\bt ^TK_{\bt }\bt )}
\eq
Where
$$
\al ^T=( \al _1,\al _2),\  \
\bt ^T=( \bt _1,\bt _2).
$$
Here and in future gaussian integrals will be written up to
multiplicative factor.\\
The Langevin equations  (\ref{f5},\ref{f6}) are linear
and after time $t$ gaussian state remains gaussian.
It is convinient to calculate (\ref{f19},\ref{f18})
in the formalism of Wigner function.
The Wigner function of  general gaussian state can be represented in the
following form
as function of means and square-means of this state
\bq
\label{f20a}
W(\RE \ga ,\IM \ga )\sim
\exp (-\frac{(\RE \ga -\lan \RE a\ran )^2}{2}K_{11}
-\frac{(\IM \ga -\lan \IM a\ran )^2}{2}K_{22}\nonumber \\
-(\RE \ga -\lan \RE a\ran )(\IM \ga -\lan \IM a\ran )K_{12})
\eq
Where
\be
\label{f21}
\left (\begin{array}{cc}K_{11}&K_{12}\\K_{12}&K_{22}
\end{array}\right )^{-1}
=\left (\begin{array}{cc}c_{11}&c_{12}\\c_{12}&c_{22},
\end{array}\right )
\ee
\bq
\label{f22}
&  &c_{11}=\lan ({\rm Re}a)^{2}\ran -\lan (
{\rm Re}a)\ran ^2\nonumber \\
&  &c_{22}=\lan ({\rm Im} a)^{2}\ran -\lan ({\rm Im}
 a)\ran ^2\nonumber \\
&  &c_{12}=c_{21}=\frac{\lan {\rm Re} a {\rm Im} a+{\rm Im}
a {\rm Re} a\ran }{2}-
\lan {\rm Im} a\ran \lan {\rm Re} a\ran
\eq
As we see this channel is gaussian.
Capacity region for two-terminal gaussian channel can be obtained
exactly.
If vectors  $\al $, $\bt $ with distribution  (\ref{f20})
are initial messages for the
channel then we have for output vector
$\ga $
\be
\label{dd1}
\ga =A_1\al +A_2\bt +z.
\ee
Where $A_1$, $A_2$ are some matrices (we shall use the specific form of these 
matrices for homodyne and heterodyne measurements ), and $z$ is
additive gaussian noisy vector
with correlation matrix $K^{-1}$
\be
\label{dd2}
\lan zz^T\ran =K^{-1}
\ee
In this case we get
\bq
\label{dd3}
&& I(\ga :\al \otimes \bt )=\ln {\rm det}(1+KA_1K_{\al
}^{-1}A_1^T+KA_2K_{\bt }^{-1}A_2^T)\nonumber \\
&& I(\ga :\al/ \bt )=\ln {\rm det}(1+KA_1K_{\al
}^{-1}A_1^T)\nonumber \\
&& I(\ga :\bt /\al )=\ln {\rm det}(1+KA_2K_{\bt }^{-1}A_2^T)
\eq
After some  not hard but tedious calculations
 we come to the concrete results.
The first case is heterodyne measurement of the mode
1 at the moment $t$ with
\be
\label{dd12}
r_1=0,\  \ r_2=0.
\ee
In this case we should choose the following initial distributions
\be
\label{dd13}
K_{\al }^{-1}=\left (\begin{array}{cc}x&0\\0&x
\end{array}\right ),\  \
K_{\bt }^{-1}=\left (\begin{array}{cc}y&0\\0&y
\end{array}\right )
\ee
With
\be
\label{dd14}
x=\half \bar{n}_1,\  \ y=\half \bar{n}_2.
\ee
Where $\bar{n}_1$, $\bar{n}_2$ are mean photon number of the
initial distributions.
In this case we have from general formulas
\bq
\label{dd15}
&& I(\ga :\al \otimes \bt )=\ln \left (1+
\frac{\bar{n}_1e^{-\ga t}+2(\bar{n}_2-\bar{n}_1)\ep (1-\ep )e^{-\ga
t}(1-\cos(t(\la _1-\la _2)))}{\half (e^{-\ga t}+\Psi +1)}\right
) \nonumber \\
&& I(\ga :\al/ \bt )=\ln \left(1+
\frac{\bar{n}_1e^{-\ga t}(1-2\ep (1-\ep )(1-\cos(t(\la _1-\la _2)))}
{\half (e^{-\ga t}+\Psi +1)}
\right )
\eq
And for $\Psi $ we have
\be
\label{dd9}
\Psi =(1-e^{-\ga t})(2\ep \bar{n}_T(\la _1)+
2(1-\ep ) \bar{n}_T(\la _2)+1)
\ee
We don't write expression for  $I(\bt :\ga /\al )$
if written formula are sufficient
for understanding the behavior of this quantity.\\
For $t=0$ these information measures coincide with well known
expression for capacity of coherent state channel.
For large  $t$ (\ref{dd15}) tend to zero.
But as we see the decay is not monotonic: the additional
oscillation
occur due to interaction between different modes.
If we take $\ep =1$ we come to one-terminal channel with
gaussian noise.
In this case capacity capacity monotonically tends
to zero.\\
Now about homodyning with arbitrary $r$.
In this case we should choose as initial distributions
\be
\label{dd16}
K_{\al }^{-1}=\left (\begin{array}{cc}x&0\\0&0
\end{array}\right ),\  \
K_{\bt }^{-1}=\left (\begin{array}{cc}y&0\\0&0
\end{array}\right )
\ee
It is product of gaussian distribution for measuring component
and delta-function for other component.
With the choice of (\ref{dd16}) we have
optimal distribution of input energy.
In this case we have the following connection between
dispersion of the distribution and mean photon number \cite{cavesdr}
\be
\label{dd17}
x=\bar{n}_1-{\rm sh}^2r_1,\  \
y=\bar{n}_2-{\rm sh}^2r_2.
\ee
In the first we consider one-terminal communication
with  arbitrary squeezing parameter $r$.
In this case we have
\be
\label{dd18}
I(\al :\ga _1 )=\half \ln \left (1+
\frac{\cos^2(\la t)(\bar{n}-{\rm sh}^2r)}
{\quart e^{-2r}+2\sin^2(\la t){\rm sh}(2r)+C }\right )
\ee
For given $\bar{n}$ optimal squeezing parameter is determined by
the following formula
\be
\label{tyy1}
e^{2r}=\frac{\sqrt{\cos^4(\la t)
+C^2+
4C(2\bar{n}+1)\cos^2(\la t)}
-\cos^2(\la t)}{C}
\ee
Where
\be
\label{tyy2}
C=
(e^{\ga t}-1)(2\bar{n}_T+1)
\ee
As we see optimal time-dependent squeezing parameter tends to zero
for large $t$.
It is well known that for zero $t$ squeezed states are more
effective than coherent one
\cite{cavesdr}.
Indeed we have
\be
\label{tyy3}
I=\ln (1+\bar{n})
\ee
for coherent states and
\be
\label{tyy4}
I=\ln (1+2\bar{n})
\ee
for squeezed states.
As we see noise beat usefulness of squeezed states, and
for optimal squeezing parameter we have (\ref{tyy1}).\\
Now about general case for homodyne measurement.
We get
\be
\label{mozi1}
I(\al :\ga _1 /\bt )=
\half \ln \left (
1+\frac{(\bar{n}_1-{\rm sh}^2r_1)u_1^2}
{\quart (u_1^2e^{-2r_1}+u_2^2e^{2r_1})
+\quart (v_1^2e^{-2r_2}+v_2^2e^{2r_2})
+\quart \Psi}\right )
\ee
\be
\label{mozi2}
I(\al \otimes \bt :\ga _1 )=
\half \ln \left (
1+\frac{(\bar{n}_1-{\rm sh}^2r_1)u_1^2+ (\bar{n}_2-{\rm
sh}^2r_2)v_1^2+ }
{\quart (u_1^2e^{-2r_1}+u_2^2e^{2r_1})
+\quart (v_1^2e^{-2r_2}+v_2^2e^{2r_2})
+\quart \Psi}\right )
\ee
Where
\bq
\label{dd5}
&&u_1=e^{-\ga t/2}(\ep \cos{(\la _1t)}+(1-\ep )\cos{(\la _2t)})
\nonumber \\
&&u_2=-e^{-\ga t/2}(\ep \sin{(\la _1t)}+(1-\ep )\sin{(\la _2t)})
\nonumber \\
&&v_1=-e^{-\ga t/2}\sqrt{\ep (1-\ep )}(\cos{(\la _1t)}-\cos{(\la _2t)})
\nonumber \\
&&v_2=e^{-\ga t/2}\sqrt{\ep (1-\ep )}(\sin{(\la _1t)}-\sin{(\la _2t)})
\eq
We can maximize by $r_1$, $r_2$
the information measure $I(\al :\ga /\bt )$
(as we know it is information transmitted by user 1),
and after this information measure
$I(\al \otimes \bt :\ga)-I(\al :\ga /\bt)$
for user 2.
The analysis show that situation with one-terminal channel is
conserved: there are optimal $r_1$, $r_2$
which tend to zero for large $t$.
The second mode introduce additional source for noise.\\
We have considered the noisy binary-access quantum-mechanical
channel, and compute capacities of this channel.
The general theorems are proved which connect capacities of the channel with 
some statistic-mechanical functions.
It is shown that second user acts as noise and can significantly
reduce the capacity of the first user. After this the information transfer by coherent and squeezed 
states  of an electromagnetic field is discussed.
Squeezed states loss their optimality under action of noise and
optimal squeezing parameter tends to zero when time tends to infinity.

\end{document}